\documentclass[prb,twocolumn,showpacs]{revtex4}
\usepackage[dvips]{graphicx}
\usepackage{latexsym}
\usepackage{amsmath}
\usepackage{amsfonts}
\usepackage{amssymb}
\usepackage{bm}

\begin{document}
\newcommand{\fig}[2]{\includegraphics[width=#1]{#2}}
\newcommand{{\vhf}}{\chi^\text{v}_f}
\newcommand{{\vhd}}{\chi^\text{v}_d}
\newcommand{{\vpd}}{\Delta^\text{v}_d}
\newcommand{{\ved}}{\epsilon^\text{v}_d}
\newcommand{{\vved}}{\varepsilon^\text{v}_d}
\newcommand{{\up}}{\uparrow}
\newcommand{{\down}}{\downarrow}
\newcommand{{\bk}}{{\bf k}}
\newcommand{{\bq}}{{\bf q}}
\newcommand{{\tr}}{{\rm tr}}
\newcommand{\pprl}{Phys. Rev. Lett. \ }
\newcommand{\pprb}{Phys. Rev. {B}}

\title{Dynamical slave-boson mean-field study of the Mott transition\\ in the Hubbard model in the large-$\textbf{\textit{z}}$ limit}

\author{Sen Zhou$^{1,2}$, Long Liang$^3$, and Ziqiang Wang$^{4}$}
\affiliation{$^1$ CAS Key Laboratory of Theoretical Physics, Institute of Theoretical Physics, Chinese Academy of Sciences, Beijing 100190, China}
\affiliation{$^2$ School of Physical Sciences \& CAS Center for Excellence in Topological Quantum Computation, University of Chinese Academy of Sciences, Beijing 100049, China}
\affiliation{$^3$ COMP Centre of Excellence, Department of Applied Physics, Aalto University School of Science, FI-00076 Aalto, Finland}
\affiliation{$^4$ Department of Physics, Boston College, Chestnut Hill, MA 02467, USA}

\date{\today}
\begin{abstract}

The Mott metal-insulator transition in the Hubbard model is studied by constructing a dynamical slave-boson mean-field theory in the limit of large lattice coordination number $z$ that incorporates the binding between doubly occupied (doublon) and empty (holon) sites.
On the Mott insulating side where all doublons and holons bond in real space into excitonic pairs leading to the charge gap, the theory simplifies considerably to leading order in $1/\sqrt{z}$, and becomes exact on the infinite-$z$ Bethe lattice.
An asymptotic solution is obtained for a continuous Mott transition associated with the closing of the charge gap at a critical value of the Hubbard $U_c$ and the corresponding doublon density $n_d^c$, hopping $\chi_d^c$ and doublon-holon pairing $\Delta_d^c$ amplitudes.
We find $U_c=U_{\rm BR} [1 -2n_d^c -\sqrt{z} (\chi_d^c +\Delta_d^c))] \simeq0.8U_{\rm BR}$, where $U_{\rm BR}$ is the critical value for the Brinkman-Rice transition in the Gutzwiller approximation captured in the static mean-field solution of the slave-boson formulation of Kotliar and Ruckenstein.
Thus, the Mott transition can be viewed as the quantum correction to the Brinkman-Rice transition due to doublon-holon binding. Quantitative comparisons are made to the results of the dynamical mean-field theory, showing good agreement.
In the absence of magnetic order, the Mott insulator is a $U(1)$ quantum spin liquid with nonzero intersite spinon hopping that survives the large-$z$ limit and lifts the $2^N$-fold degeneracy of the local moments.
We show that the spinons are coupled to the doublons/holons by a dissipative compact $U(1)$ gauge field in the deconfined phase, realizing the spin-charge separated gapless spin liquid Mott insulator.

\typeout{polish abstract}
\end{abstract}

\pacs{71.10.-w, 71.10.Fd, 71.27.+a, 74.70.-b}

\maketitle

\section{Introduction}
A Mott insulator is a fundamental quantum electronic state driven by large Coulomb repulsion \cite{mottpaper, mottrmp, mottbook}.
It is protected by a nonzero energy gap for charge excitations, but not associated with any symmetry breaking.
A Mott insulator differs from the other class of correlation-driven insulators (e.g., magnets), better termed as Landau insulators, whose origins require symmetry breaking order parameters produced by the residual quasiparticle (QP) interactions from a parent Fermi liquid state.
The most striking feature of a Mott insulator is the separation of charge and spin degrees of freedom of an electron that completely destroys the coherent QP excitations.
A ubiquitous example of Mott insulator is the quantum spin liquid (QSL) where the spins are short-range correlated but do not exhibit any symmetry-breaking long-range order \cite{anderson73, wen91, palee08, balents10}.
Overwhelming evidence for QSLs has been observed in the $\kappa$-organics {\it near} the Mott metal-insulator transition \cite{kanoda03, kanoda05, kanoda08, matsuda08} and in frustrated quantum magnets \cite{han2012, li2015, shen2016} that are deep in the Mott insulating state.

The Mott insulator and the Mott transition are at the heart of the strong correlation physics since it is conceivable that the Mott insulator is the ultimate parent phase of strong correlation from which many novel quantum states can emerge.
Indeed, strong correlation often results in an insulating ground state with antiferromagnetic long-range order, where the low-energy physics is described by the Heisenberg type of spin models, which can be viewed as instabilities of the spin liquids due to the condensation of low energy spin excitations in the Mott insulator.
In addition to QSLs and magnetic ordered states, doping a Mott insulator can lead to the pseudogap phenomenon and unconventional high-$T_c$ superconductivity \cite{anderson87, kivelson, imadarmp, phillips, leermp, weng}.

The prototypical model for the Mott physics is the half-filled single-band Hubbard model with purely on-site Coulomb repulsion $U$.
The Hilbert space is thus a product of the local Hilbert space on a single lattice site that consists of the doubly occupied (doublon), empty (holon), and singly occupied (spinon) states.
The excitonic binding between the oppositely charged doublons (D) and holons (H) is believed to play an essential role in describing the Mott insulator and the Mott transition in strongly correlated Mott-Hubbard systems.
This idea was advocated sometime ago \cite{mottpaper, kohn, kemeny, kaplan} and studied in the context of improved variational Gutzwiller wave functions \cite{yokoyama, capello}.
More recently, the idea has been made more explicit in the field theory description \cite{leigh}, improved saddle-point solution \cite{zhouwangwang} of the Kotliar-Ruckenstein slave-boson functional integral formulation of the Hubbard model \cite{kr}, and other numerical \cite{sato, mckenzie} approaches.
The static saddle-point solution of the slave-boson path integral \cite{kr} correctly captures the Gutzwiller approximation \cite{gutzwiller} and gives rise to the Brinkman-Rice (BR) metal-insulator transition at $U_{\rm BR}$ \cite{kr,dv-rmp84} where the renormalized mass of the QPs diverges and the band becomes flat. This is, however, not a rather crude approximation of the Mott transition, since the interactions between the doublons and holons as well as the incoherent excitations have been ignored. In an effort to go beyond the Gutzwiller approximation and the BR picture, an improved saddle point solution beyond the static limit was constructed in Ref. \cite{zhouwangwang} on two-dimensional bipartite lattices.
It was elucidated that the doublon-holon (D-H) binding governs the incoherent excitations and plays a key role in the Mott transition.
On the Mott insulator side at large $U$, although the D/H condensate vanishes, together with the disappearance of the coherent QPs, the D/H density remains nonzero, but with all the doublons bond to the holons.
With decreasing $U$, the D/H density increases and D-H binding energy decreases.
At a critical value $U_c$, the D-H excitation gap closes and a D/H single-particle condensate starts to develop, marking the onset of the Mott transition.
Despite the success in capturing the essential Mott physics, the improved dynamical saddle-point solution is uncontrolled and quantitatively unreliable.

In this work, we construct a controlled dynamical slave-boson mean-field (SBMF) theory that incorporates the D-H binding and becomes exact in the limit of large coordination number $z$ (Section II).
The dynamical SBMF theory is also referred to as the D-H binding theory in the rest of the paper.
It turns out that the theory simplifies considerably on the Mott insulating side to leading order in $1/\sqrt{z}$ and becomes exact in the large-$z$ limit. We therefore study the Mott transition from the large-$U$ Mott insulating side (Section III.A).
The asymptotic solution obtained on the infinite-$z$ Bethe lattice
exhibits a continuous Mott transition from an insulating QSL to a correlated metal, where the closing of the Mott gap and the onset of the QP coherence coincide at the same $U_c$.
We demonstrate that in the presence of D-H binding captured by the dynamical SBMF theory, the BR transition is preempted by the Mott transition since $U_c<U_{\rm BR}$, and the Mott insulator is characterized by the incoherent upper and lower Hubbard bands separated by the Mott gap.
A key feature of the asymptotic solution is that on the insulating side of the Mott transition, quantum spin fluctuations via the intersite spinon correlations remains and survives in the large-$z$ limit.
Various physical quantities are calculated from simple and transparent expressions analytically at the transition point $U_c$, as well as to leading order in the Mott insulator at $U>U_c$.
The results are quantitatively compared to and found to agree well with those obtained from the dynamical mean-field theory\cite{gk92} (DMFT) with various numerical quantum impurity solvers, which is exact in the large-$z$ limit.
In Section III.B, we derive the effective action for the compact gauge field in the large-$z$ limit and show that the emergent dissipative dynamics drives the gauge field to the deconfinement phase where the spin-charge separated U(1) spin liquid is stable. The summary and discussions are presented in Section IV.

\section{Model and large-$\textbf{\textit{z}}$ theory}
We start with the half-filled Hubbard model given by
\begin{equation}
H = -{t\over \sqrt{z}} \sum_{\langle ij \rangle,\sigma} c_{i\sigma }^ \dagger  c_{j\sigma }  + {\rm h.c.} + U\sum_i {n_{i \uparrow } n_{i \downarrow } },
\label{h}
\end{equation}
where the $t$-term describes electron hopping on a lattice with $z$ nearest neighbor (NN) bonds, and the $U$-term is the on-site Coulomb repulsion.
When the quantum states are spatially extended, the NN single-particle correlator scales with the coordination number as $\langle c_{i\sigma}^\dagger c_{j\sigma}\rangle\sim 1/\sqrt{z}$.
As a result, the ${1/ \sqrt{z}}$-scaling for the hopping $t$ in Eq.~(\ref{h}) is necessary in order to maintain a finite kinetic energy in the large-$z$ limit \cite{metzner89}.
This rescaling is used in the DMFT \cite{gk92}.

To construct a strong-coupling theory that is nonperturbative in $U$, Kotliar and Ruckenstein \cite{kr} introduced a spin-1/2 fermion $f_\sigma$ and four slave bosons $e$ (holon), $d$ (doublon), and $p_{\sigma}$ to represent the local Hilbert space for the empty, doubly-occupied, and singly occupied sites respectively: $\vert0 \rangle= e^\dagger \vert \text{vac} \rangle$,  $|\!\!\! \uparrow \downarrow \rangle =d^\dagger f_\downarrow^\dagger f_\uparrow^\dagger \vert \text{vac} \rangle$, and $\vert \sigma\rangle= p_\sigma^\dagger f_{\sigma}^\dagger \vert \text{vac}\rangle$.
The physical Hilbert space is obtained under the local constraints for the completeness
\begin{equation}
e_i^\dagger e_i + \sum_\sigma p_{i\sigma}^\dagger p_{i\sigma} +d_i^\dagger d_i =1,
\label{completeness}
\end{equation}
and the consistency
\begin{equation}
f_{i\sigma}^\dagger f_{i\sigma} =p_{i\sigma}^\dagger p_{i\sigma} +d_i^\dagger d_i.
\label{consistency}
\end{equation}
The Hubbard model is thus faithfully represented by
\begin{equation}
 {H}=-{t\over \sqrt{z}} \sum_{\langle ij \rangle,\sigma} Z_{i\sigma}^\dagger Z_{j\sigma} f_{i\sigma}^\dagger f_{j\sigma} + {\rm h.c.} + U\sum_i d_i^\dagger d_i,
\label{hslaveboson}
\end{equation}
where the composite bosonic operator
\begin{equation}
Z_{i\sigma}=L_{i\sigma}^{-1/2}(p_{i\bar\sigma}^\dagger d_i+e_i^\dagger p_{i\sigma})R_{i\bar\sigma}^{-1/2}.
\label{zboson}
\end{equation}
The operators $L_{i\sigma}=1 -d_i^\dagger d_i-p_{i\sigma}^\dagger p_{i\sigma}$ and $R_{i\bar\sigma}=1-e_i^\dagger e_i -p_{i\bar\sigma}^\dagger p_{i\bar\sigma}$ should be understood as projection operators for hardcore bosons with unit eigenvalues, and the choice of the $-1/2$ power in Eq.~(\ref{zboson}) reproduces the Gutzwiller approximation at the level of the static saddle point \cite{kr}.

Unlike fermions that subject to Pauli exclusion principle, the bosons have a remarkable property: a macroscopically large number of them can condense into a single quantum state.
Thus the $Z_{i\sigma}$-boson can be decomposed into a single-particle condensate part $Z_{i\sigma,0}$ and an uncondensed ``normal'' or fluctuating part $\tilde{Z}_{i\sigma}$, i.e., $Z^{(\dagger)}_{i\sigma} =Z_{i\sigma,0} +\tilde{Z}^{(\dagger)}_{i\sigma}$, and similarly for the slave bosons.
Consequently, the single-particle correlator of $Z$-bosons has two contributions
\begin{equation}
\langle Z^\dagger_{i\sigma} Z_{j\sigma} \rangle =Z_{i\sigma,0} Z_{j\sigma,0} +\langle \tilde{Z}^\dagger_{i\sigma} \tilde{Z}_{j\sigma} \rangle,
\label{Zcorrelator}
\end{equation}
where the first term comes from the single-boson condensate and does not scale with the coordination number $z$ or the distance
$r_{ij}=\vert {\bf r}_i -{\bf r}_j \vert$ between site $i$ and $j$. The second term $\langle \tilde{Z}^\dagger_{i\sigma} \tilde{Z}_{j\sigma} \rangle$ comes from the uncondensed fluctuating bosons and decreases with increasing $z$ and $r_{ij}$ according to $z^{-r_{ij}/2}$.
More explicitly, on the NN bonds, $Z_{i\sigma,0} Z_{j\sigma,0} \sim 1$, whereas $\langle \tilde{Z} ^\dagger _{i\sigma} \tilde{Z}_{j\sigma} \rangle\sim1/\sqrt{z}$.
As a result, the kinetic hopping energy of uncondensed bosons is of higher order by $1/\sqrt{z}$ than that of the condensed bosons, and thus negligible in the large-$z$ limit and the Hamiltonian in Eq. (\ref{hslaveboson}) reduces to the static saddle point solution characterized by the single-particle condensation of all the slave bosons \cite{kr}.
Thus, taking the large-$z$ limit this way results in the static SBMF theory which is equivalent to the Gutzwiller approximation.
The BR metal-insulator transition takes place when the condensate density of $Z$-boson is driven zero by the vanishing of the D/H density for $U>U_{\rm BR}$, which is equivalent to having a divergent effective mass for the QPs.

In order to include the effects of the fluctuating $Z$-bosons, it is necessary to treat the contributions from the condensed and uncondensed bosons on equal footing.
As pointed out explicitly in the formulation of the \textit{bosonic} DMFT\cite{bdmft}, this can be achieved by different rescalings of the bosonic hopping amplitudes for the condensate and the fluctuating parts.
Utilizing this rescaling for the bosons, the slave-boson formulation of the Hubbard model in Eq. (\ref{hslaveboson}) is rewritten as
\begin{align}
 {H}=&-{t\over\sqrt{z}}\sum_{\langle ij \rangle,\sigma} \left( Z_{i\sigma,0} Z_{j\sigma,0} +\sqrt{z}\tilde{Z}_{i\sigma}^\dagger \tilde{Z}_{j\sigma} \right) f_{i\sigma}^\dagger f_{j\sigma} \nonumber \\
&+ {\rm h.c.} + U\sum_i d_i^\dagger d_i.
\label{hslaveboson2}
\end{align}
Since the NN correlator for the fluctuating bosons $\langle \tilde{Z} ^\dagger _{i\sigma} \tilde{Z}_{j\sigma} \rangle\sim1/\sqrt{z}$, it contributes to the hopping integral on equal footing as the condensed part.
On account of the fermion correlator $\langle f^\dagger_{i\sigma} f_{j\sigma} \rangle\sim 1/\sqrt{z}$, this lead to a finite kinetic energy coming from the uncondensed and fluctuating bosons beyond the Gutzwiller approximation or the static SBMF theory in the large-$z$ limit.

The rescaled Hamiltonian Eq.~(\ref{hslaveboson2}) has a remarkably property that it simplifies considerably on the Mott insulator side where the D/H single-particle condensate vanishes, i.e. $d_{i0} = e_{i0} =0$, which implies $Z_{i0}=0$ by Eq.~(\ref{zboson}).
As a result, the kinetic energy solely comes from the uncondensed bosons accompanied by the back-flow of the fermions between the neighboring sites, which is a signature of electron fractionalization in the Mott insulating state.
To see that the electrons must be incoherent, it is instructive to note that since the correlators of the $f$-fermion and the $Z$-bosons both scale as $1/\sqrt{z}$, the electron intersite correlator $\langle c_{i\sigma}^\dagger c_{j\sigma}\rangle\sim 1/z$, which is completely different from that of the coherent QP hopping behavior on the metallic side, yet contributes to a finite kinetic energy in this large $z$-limit.

We thus study the Mott transition from the Mott insulating side at large $U>U_c$.
The absence of D/H condensate leads to $\tilde{Z}_{i\sigma} =L_{i\sigma}^{-1/2} (p_{i\bar\sigma,0} {d}_i +{e}_i^\dagger p_{i\sigma,0}  ) R_{i\bar\sigma}^{-1/2}$, where $d_i$ and $e_i$ are the fluctuating D/H having a nonzero density $n_d=n_e=\langle d_i^\dagger d_i\rangle=\langle e_i^\dagger e_i\rangle\neq 0$.
The $p_{i\sigma}$ bosons representing single-particle occupation condense into $c$-numbers with $p_{i\sigma,0} =p_{i\sigma,0}^\dagger =p_0$ in the absence of magnetism.
Furthermore, the operators $L_{i\sigma}$ and $R_{i\sigma}$ contained in $\tilde{Z}_{i\sigma}$ should not introduce additional intersite correlations to leading order in $1/\sqrt{z}$, in contrast to uncontrolled saddle point approximations with D-H binding in two dimensions \cite{zhouwangwang}.
They can thus be written in terms of the {\it local} densities, leading to $L_{i\sigma}=R_{i\sigma}=1/2$ at half-filling.
Hence, on the Mott insulating side,
\begin{equation}
\tilde{Z}_{i\sigma}=2p_0(d_i+e_i^\dagger),
\label{zboson-largez}
\end{equation}
to leading order in $1/\sqrt{z}$ and the Hamiltonian in Eq.~(\ref{hslaveboson2}) becomes
\begin{eqnarray}
H=&-&4p_0^2t\sum_{\langle ij \rangle}\bigl[(d_i^\dagger d_j+e_j^\dagger e_i+e_i d_j +d_i^\dagger e_j^\dagger ) f_{i\sigma}^\dagger f_{j\sigma}
\nonumber \\
&+&{\rm h.c.}\bigr]
+ U\sum_i d_i^\dagger d_i.
\label{h-largez}
\end{eqnarray}

It is straightforward to write down the path integral of the model.
The condensation of the $p_\sigma$ bosons collapses two of the operator constraints in Eqs~(\ref{completeness}-\ref{consistency}) into consistency equations for particle densities $n_d+ p_0^2 =n_\sigma^f$.
The remaining one can be written as
\begin{equation}
e_i^\dagger e_i-d_i^\dagger d_i+\sum_\sigma f_{i\sigma}^\dagger f_{i\sigma}=1,
\label{constraint}
\end{equation}
which corresponds to the unbroken internal $U(1)$ gauge symmetry and specifies the gauge charges of the particles.
Eq.~(\ref{constraint}) shows that increasing the spinon number by one must be accompanied by either destroying a holon or creating a doublon at the same site.
The partition function can be written down as an imaginary-time path integral
\begin{equation}
Z=\int {\cal D}[f^\dagger,f]{\cal D}[d^\dagger,d]{\cal D}[e^\dagger e]{\cal D}[a_0,a] {\cal D}\lambda e^{-\int_0^\beta {\cal L} d\tau},
\label{partition}
\end{equation}
with the Lagrangian
\begin{eqnarray}
{\cal L}&=&\sum_i \bigl[
d_i^\dagger(\partial_\tau -ia_0)d_i+e_i^\dagger(\partial_\tau
+ ia_0)e_i\nonumber\\
&+& f_{i\sigma}^\dagger(\partial_\tau+ia_0)f_{i\sigma}\bigr]-H_f-H_b
\label{lagrangian} \\
&+&i\sum_i\lambda_i(d_i^\dagger d_i+ e_i^\dagger e_i+2p_0^2-1),
\nonumber
\end{eqnarray}
where $\lambda_i$ is a Langrange multiplier.
The decoupled fermion and boson Hamiltonian \cite{leermp} are given by
\begin{eqnarray}
H_f=&-&{t_f\over\sqrt{z}}\sum_{\langle i,j\rangle}(e^{ia_{ij}}f_{i\sigma}^\dagger f_{j\sigma}+{\rm h.c.})
\label{hf} \\
H_b=&-&{t_b\over\sqrt{z}}\sum_{\langle i,j\rangle}\bigl[e^{-ia_{ij}}(e_j^\dagger e_i+d_i^\dagger d_j
\label{hb} \\
&+&e_id_j+d_i^\dagger e_j^\dagger)+{\rm h.c.}\bigr]+{U\over2}\sum_i(d_i^\dagger d_i+e_i^\dagger e_i),
\nonumber
\end{eqnarray}
with
\begin{equation}
t_f=8tp_0^2\sqrt{z}(\chi_d+\Delta_d), \quad t_b=8tp_0^2\sqrt{z}\chi_f.
\label{tbf}
\end{equation}
In a stationary state, $\chi_d=\langle d_i^\dagger d_j\rangle=\langle e_j^\dagger e_i\rangle$ is the quantum average of the D/H nearest neighbor hopping, $\chi_f =\langle f_{i\sigma}^\dagger f_{j\sigma}\rangle$ the spinon hopping per spin, and $\Delta_d=\langle d_i^\dagger e_j^\dagger\rangle=\langle e_id_j\rangle$ is the D-H binding order parameter.
In Eqs~(\ref{lagrangian}-\ref{hb}), the spinons and the D/H are coupled by the emergent $U(1)$ gauge fields $a_0$ and $a_{ij}$ associated with the constraint in Eq.~(\ref{constraint}).
Physically, the instantons of this compact gauge field correspond to the tunneling events where the spinons and D/H tunnel in and out of the lattice sites \cite{ioffelarkin}.

\section{Mott transition and spin liquid Mott insulator}

\subsection{Asymptotic solution for Mott transition}
We will first obtain the stationary state solution with $a_0 =a_{ij} =0$, and then study the properties of the gauge field fluctuations.
Eq.~(\ref{hf}) shows that the spinon hopping amplitude is $t_f/ \sqrt{z}$ where $t_f$ defined in Eq.~(\ref{tbf}) is proportional to the D/H intersite correlations.
We will show that the latter leads to a renormalized narrow spinon band with a bandwidth on the order of the exchange coupling $J\sim t^2/U$ in the large $U$ limit.
The spinon kinetic energy per site is $K_f=(t_f/ t)K_0$ where $K_0 =2\int_0^D \rho_0(\omega) \omega d \omega$ is that for noninteracting electrons with hopping $t/\sqrt{z}$ and $\rho_0$ is the corresponding semicircle density of states $\rho_0(\omega)={2\over \pi D} \sqrt{1 -(\omega/D)^2}$ on the infinite-$z$ Bethe lattice \cite{dmftrmp96} with a half-bandwidth $D=2t$.
Note that both $D$ and $t$ are order one quantities, since the $1/\sqrt{z}$ factors in Eqs~(\ref{hf}) and (\ref{hb}) are dynamically generated by the NN intersite correlators $(\chi_d, \Delta_d, \chi_f)\sim 1/\sqrt{z}$ in Eq.~(\ref{tbf}) where $t_b$ and $t_f$ are of order one.
Thus, $K_0=8t/3\pi = 4D/3\pi$ and $K_f=8t_f/3\pi$.
Since $K_f$ can also be written as $K_f=4t_f\sqrt{z}\chi_f$, we obtain readily $\chi_f={1\over\sqrt{z}}{2\over3\pi}$, independent of $U$.
The boson hopping parameter $t_b$ in Eq.~(\ref{tbf}) is thus given by $t_b =16 p_0^2 t /3 \pi$, which is on the order of $t$.
Hence, the spectrum of charge excitations residing in the D/H sector has a bandwidth on the order of the bare electron bandwidth, giving rise to the broad incoherent spectral weight induced by strong correlation.

From Eqs (\ref{lagrangian}) and (\ref{hb}), the stationary state bosonic Hamiltonian in the D/H sector is
\begin{equation}
H_{\rm D/H} \!=\! \int_{-D}^D \!\! d\omega\rho_0(\omega) \left[ d_\omega^\dagger,e_\omega \right] \!\! \left[
\begin{array}{cc} \varepsilon_\omega & - \Delta_\omega \\
-\Delta_\omega & \varepsilon_\omega \end{array} \right] \!\!
\left[\begin{array}{c} d_\omega \\ e_\omega^\dagger
\end{array} \right]\!\!, \label{hdh}
\end{equation}
where $\varepsilon_\omega={U\over2}+\lambda-{t_b\over t}\omega$, $\Delta_\omega ={t_b\over t} \omega$ are the D/H kinetic and pairing energies; $\lambda=\langle i\lambda\rangle$.
Diagonalizing $H_{\rm D/H}$ by Bogoliubov transformation produces two degenerate branches for the D/H excitations,
\begin{equation}
\Omega_\omega=\sqrt{\varepsilon_\omega^2-\Delta_\omega^2}.
\label{bosonomega}
\end{equation}
The Mott insulator is thus an excitonic insulator and the Mott gap is given by the charge gap in $\Omega_\omega$,
\begin{equation}
G_{\rm Mott}(U)=2\Omega_D=2\sqrt{\left({U\over2}+\lambda\right)
\left({U\over2}+\lambda-4t_b\right)}.
\label{mottgap}
\end{equation}
The physical condition for a real $\Omega$ requires $U\ge 8t_b -2\lambda$ and the equal sign determines the critical $U_c$ for the Mott transition where $G_{\rm Mott}(U_c)=0$.
\begin{figure}
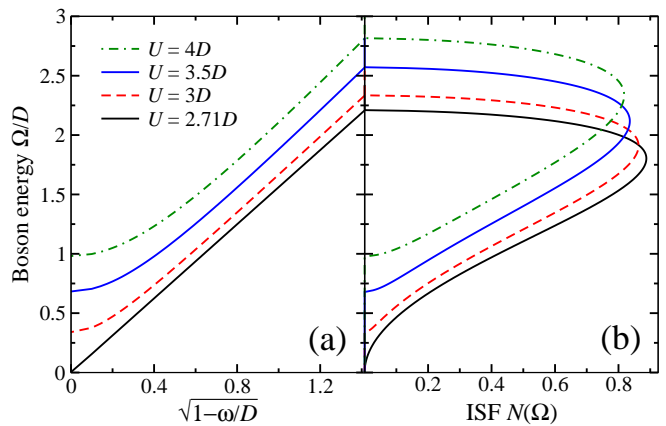

\begin{center}
\fig{3.4in}{fig1.eps}\caption{The doublon/holon energy spectrum (a) and the corresponding spectral density of states (b) for different $U$.}
\end{center}
\end{figure}

Minimizing the energy leads to the self-consistent equations, $p_0^2={1\over2}-n_d$, $\lambda=4K_0\sqrt{z}(\chi_d+\Delta_d)$, and
\begin{eqnarray}
n_d&=&{1\over2}\int_{-D}^{D}\left({\varepsilon_\omega\over\Omega_\omega}
-1\right)\rho_0(\omega)d\omega,
\label{nd} \\
\sqrt{z} \chi_d&=&{1\over 2D}\int_{-D}^D {\varepsilon_\omega\over\Omega_\omega}\omega\rho_0(\omega)d\omega,
\label{chid} \\
\sqrt{z} \Delta_d&=&{1\over 2D}\int_{-D}^D {\Delta_\omega\over\Omega_\omega}\omega\rho_0(\omega)d\omega.
\label{deltad}
\end{eqnarray}
Eq.~(\ref{nd}) shows that the nonzero D/H density is entirely due to the quantum fluctuations above the Mott gap in $\Omega_\omega$ for $U>U_c$.
Lowering $U$ toward $U_c$, $G_{\rm Mott}$ must reduce to host the increased D/H density until $G_{\rm Mott}=0$ at $U=U_c$ where the D/H condensation emerges and, as we shall shown, the continuous Mott transition takes place.

Solving these equations self-consistently, we obtain the properties of the Mott insulator and the Mott transition.
The D/H excitation spectrum is plotted in Fig.~1(a), showing the closing of the Mott gap as $U$ is reduced toward $U_c$.
Note that the calculated spectral density of states, i.e., the integrated spectral function (ISF), $N^{\rm D/H}(\Omega)$ shown in Fig.~1(b) vanishes quadratically upon gap closing, which ensures that the Mott transition is continuous at zero temperature.

\subsubsection{Critical properties at Mott transition}
Remarkably, the critical properties of the transition can be determined analytically.
First, setting the Mott gap $G_{\rm Mott}(U_c)=0$ gives
\begin{equation}
U_c=8t_b^c-2\lambda^c,
\end{equation}
where the script $c$ denotes the critical values of the corresponding quantity at the transition point.
Next, using the expressions for $t_b$ in Eq.~(\ref{tbf}) and $\lambda$ given above Eq.~(\ref{nd}), we obtain
\begin{equation}
U_c=U_{\rm BR}[1-2n_d^c-\sqrt{z}(\chi_d^c+\Delta_d^c)],
\label{uc}
\end{equation}
where $U_{\rm BR}=8K_0=32D/3\pi$ is the critical value for the BR transition on the Bethe lattice and $(n_d^c,\chi_d^c,\Delta_d^c)$ are the critical values of the doublon density, doublon hopping, and the D-H binding, respectively.
Eq.~(\ref{uc}) reveals the much desirable connection between the Mott transition and the BR transition.
It shows that the Mott transition can be viewed as the quantum correction to the BR transition due to D-H binding.
Since $U_c<U_{\rm BR}$, the BR transition is preempted by the Mott transition and unobservable in the Hubbard model.

At $U=U_c$, it is straightforward to calculate the D/H kinetic and pairing energies in Eq.~(\ref{hdh}) to obtain $(\varepsilon_\omega^c,\Delta_\omega^c)=(1-2n_d^c){8D\over3\pi}
(2-{\omega\over D},{\omega\over D})$, such that the critical D/H excitation spectrum in Eq.~(\ref{bosonomega}) becomes
\begin{equation}
\Omega_\omega^c=(1-2n_d^c){16D\over3\pi}\sqrt{1-{\omega\over D}}.
\label{omegac}
\end{equation}
The spectrum is independent of $\chi_d$ and $\Delta_d$ and agrees with the one shown in Fig.~1(a) at $U=2.71D$.
Furthermore, the ratios $\varepsilon_\omega^c /\Omega_\omega^c$ and $\Delta_\omega^c /\Omega_\omega^c$ that enter Eqs~(\ref{nd}-\ref{deltad}) are simple universal functions such that these integrals can be evaluated analytically to obtain the critical quantities at the Mott transition,
\begin{eqnarray}
n_d^c&=&{12\sqrt{2}-5\pi\over 10\pi}\simeq0.040
\label{ndc} \\
\sqrt{z}\chi_d^c&=&{2\over 35\pi}\sqrt{2}\simeq0.026
\label{chidc} \\
\sqrt{z}\Delta_d^c&=&{22\over 105\pi}\sqrt{2}\simeq0.094.
\label{deltadc}
\end{eqnarray}
Inserting these values into Eq.~(\ref{uc}), we obtain the critical Hubbard interaction for the Mott transition,
\begin{equation}
U_c\simeq0.80\cdot U_{\rm BR}\simeq2.71D,
\label{uc2}
\end{equation}
at which the charge gap closes and the QP coherence emerges with the D/H condensate simultaneously.
\begin{figure}
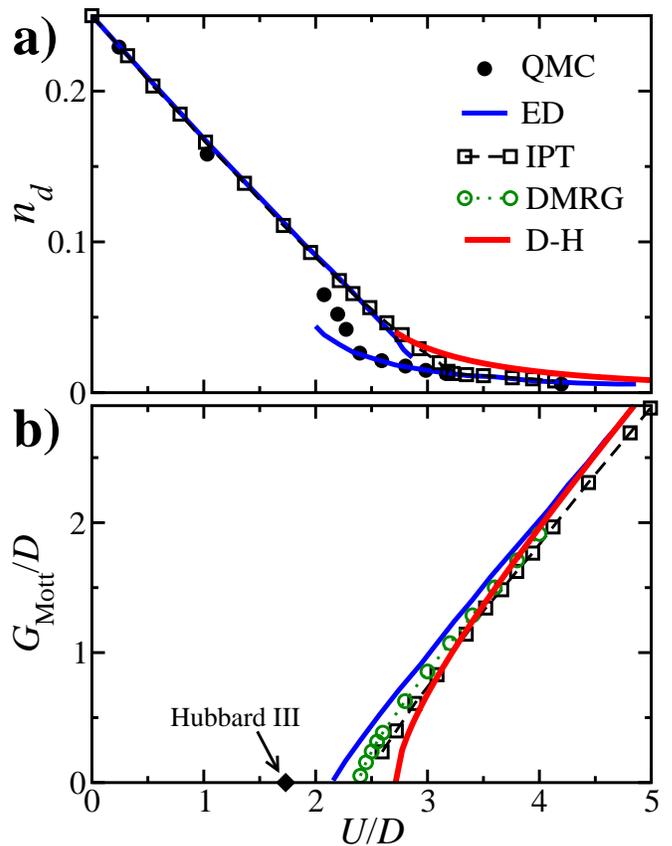

\begin{center}
\fig{3.4in}{fig2.eps}
\caption{Mott insulator in the large-$z$ limit of the D-H binding theory (red lines). (a) The doublon density as a function of $U$.
(b) The Mott gap in the charge sector as a function of $U$.
The DMFT results obtained using different impurity solvers are also shown for comparison (data from Ref.\cite{dmftrmp96,karski}): quantum monte carlo (QMC - solid black circles), exact diagonalization (ED - blue lines), iterative perturbation theory (IPT- open squares), and dynamical density matrix renormalization group (DMRG - open circles).}
\end{center}
\end{figure}

\subsubsection{Doublon density and Mott gap for $U>U_c$}

In Figs ~2(a) and 2(b), the calculated doublon density and Mott gap are plotted in red solid lines as a function of $U/D$ on the insulating side of the Mott transition.
Various single-site DMFT results \cite{dmftrmp96,karski,dv2012} are also plotted in Fig.~2 for comparison solely for the purpose of benchmarking the results in the charge sector, despite the different large-$z$ limit and the continuous Mott transition to a spin liquid at a single $U_c$.
These zero temperature ground state properties are known to be difficult to obtain reliably in the DMFT and near the Mott transition, as reflected in the discrepancies between the results obtained using different quantum impurity solvers \cite{dmftrmp96}.
Fig.~2(a) shows that the doublon density decays algebraically with increasing $U$.
Indeed, Eqs~(\ref{nd}-\ref{deltad}) can be solved analytically to obtain the large $U$ behaviors
\begin{eqnarray}
n_d &=&\left({8\over3\pi}\right)^2 {t^2\over U^2} + {\cal O}\left({t\over U}\right)^4,
\label{ndlargeu} \\
\sqrt{z}\chi_d&=&4\left({8\over3\pi}\right)^3 {t^3\over U^3} + {\cal O}\left({t\over U}\right)^5,
\label{chidlargeu} \\
\sqrt{z}\Delta_d &=&{4\over3\pi}{t\over U} + {\cal O}\left({t\over U}\right)^3.
\label{deltadlargeu}
\end{eqnarray}
In Fig.~3, the evolution of $(n_d, \sqrt{z}\chi_d, \sqrt{z}\Delta_d)$ as a function of $D/U$ is shown on a log-log plot.
It can be seen that the general self-consistent solutions of Eqs~(\ref{nd}-\ref{deltad}) represented by the solid lines merges with the corresponding dashed-lines describing the asymptotic large-$U$ behaviors given in Eqs~(\ref{ndlargeu}-\ref{deltadlargeu}).
Thus, the holons and doublons are always present at any $U$.
The binding of the opposite charges on the energy scale of the Mott gap $G_{\rm Mott}(U)$ makes it possible to treat them as localized quantum defects in the Heisenberg model description of the physics on the energy scale of the exchange coupling $J$, provided that $U$ is large enough such that $G_{\rm Mott}(U)\gg J$.
Note that $\sqrt{z}\Delta_d \gg n_d \gg \sqrt{z}\chi_d$ in the large-$U$ regime.
As a consequence, the large-$U$ physics of the spin-liquid Mott insulator is controlled by D-H binding. We will come back to the physical significance of the latter shortly.

\begin{figure}
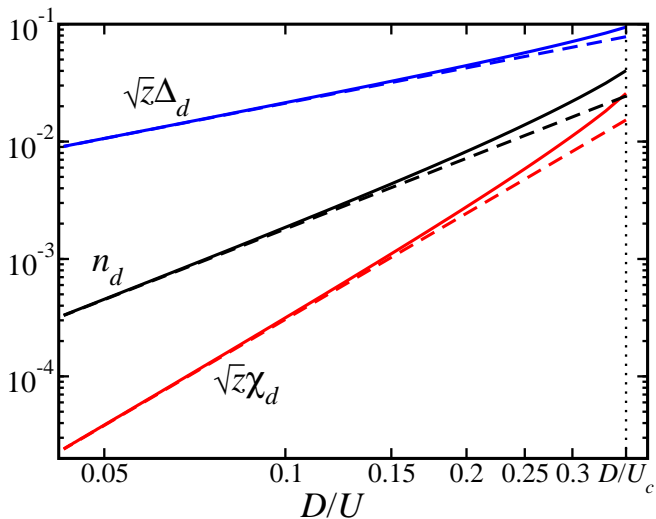

\begin{center}
\fig{3.4in}{fig3.eps}\caption{Evolution of the doublon density $n_d$, D/H hopping $\sqrt{z} \chi_d$, and D-H binding $\sqrt{z} \Delta_d$ as a function of $D/U$ on a log-log plot in the Mott insulating state. Solid lines: fully self-consistent solutions of Eqs (\ref{nd} -\ref{deltad}). Dashed lines: asymptotic solutions in the large-$U$ limit given in Eqs~(\ref{ndlargeu}-\ref{deltadlargeu}). The vertical dotted line indicates the critical $D/U_c\simeq0.37$.}
\end{center}
\end{figure}

As $U$ is reduced towards $U_c$, the calculated doublon density $n_d$ in Fig.~2(a) approaches and merges at $U_c$ smoothly with those obtained for $U<U_c$ by the DMFT using the {\it zero-temperature} iterative perturbation theory (IPT) and exact diagonalization (ED) impurity solvers.
This is reassuring since the large-$z$ limit used in the DMFT is both natural and appropriate for the Hubbarrd model on the metallic side of the Mott transition.
The critical behavior of the Mott gap near $U_c$ can also be obtained analytically from Eq.~(\ref{mottgap}), $G_{\rm Mott}(U) =\alpha\sqrt{U-U_c}$, $\alpha =2\sqrt{2t_b^c}\simeq2.61\sqrt{t}$, where the square-root singularity is clearly seen in Fig.~2(b).
The Mott gap increases with $U$ and approaches that obtained in the DMFT using ED impurity solver and exhibit the asymptotic behavior in the large-$U$ limit $G_{\rm Mott}(U\gg D)=U$ seen from Eq.~(\ref{mottgap}).

\subsubsection{Spectroscopy of spin liquid Mott insulator}
Figs ~4(a) and 4(b) show the spectroscopic properties on the Mott insulating side with comparison to the corresponding DMFT results.
They are obtained by calculating the local electron Green's function
\begin{equation}
G_\sigma(\tau)=-\langle {\rm T}_\tau c_{i\sigma}(\tau)c_{i\sigma}^\dagger(0)\rangle=G_{\sigma}^f(\tau)G_Z(\tau),
\label{gc}
\end{equation}
where $G_\sigma^f$ and $G_Z$ are the corresponding local Green's functions of the spinon and the $Z$-boson (linear combinations of the D/H).
In Matsubara frequency space, Eq.~(\ref{gc}) amounts to a convolution
\begin{equation}
G_\sigma(i\omega_n)=\sum_{i\nu_n}G_\sigma^f(i\omega_n-i\nu_n)G_Z(i\nu_n)
\label{convolution}
\end{equation}
of the spinon and the D/H local Green's functions \cite{zhouwangwang}
\begin{eqnarray}
G_\sigma^f(i\omega_n)&=&\int d\epsilon \rho_0(\epsilon)G_\sigma^f(\epsilon,i\omega_n), \\
G_Z(i\nu_n)&=&\int d\epsilon \rho_0(\epsilon)G_Z(\epsilon,i\nu_n).
\end{eqnarray}
The electron spectral density is given by
$N_\sigma(\omega)=-{1\over\pi} {\rm Im}G_\sigma(i\omega_n\to\omega+i0^+)$.
Fig.~4(a) shows $N_\sigma(\omega)$ obtained at $U=4D$, exhibiting the upper and the lower Hubbard bands separated by the Mott gap, in broad semi-quantitative agreement with the DMFT results obtained by IPT and the more recent dynamical density matrix renormalization group (DMRG) impurity solvers \cite{dmftrmp96,karski}.
The spectral density of the spinons $N_\sigma^f(\omega)$ also shown in Fig.~2(a) is, on the other hand, gapless and contributes to the thermodynamic properties of the spin liquid at low temperatures.
The spinon half bandwidth is $D_f=2t_f$ in the large-$z$ limit, where $t_f$ is the spinon hopping integral given in Eq.~(\ref{tbf}).
Let's consider the physics when $U$ is large. In this case, $t_f$ can be readily evaluated using the solutions for $(n_d, \sqrt{z}\chi_d, \sqrt{z}\Delta_d)$ obtained in Eq.~(\ref{ndlargeu}-\ref{deltadlargeu}).
Note that since $\sqrt{z}\Delta_d \propto {t/U} \gg \sqrt{z}\chi_d$ in the large-$U$ limit, it dominates the contributions to $t_f$ and leads to $t_f={4\over3\pi}{4t^2\over U}$.
Thus, the spinon hopping amplitude and bandwidth are controlled by the exchange coupling $J$, capturing the physics of the gapless $U(1)$ spin liquid phase in the effective Heisenberg model.
Moreover, the analysis shows that the origin of the exchange coupling $J$ on the insulating side of the Mott transition is intimately connected to the physics of D-H binding in the Hubbard model.
\begin{figure}
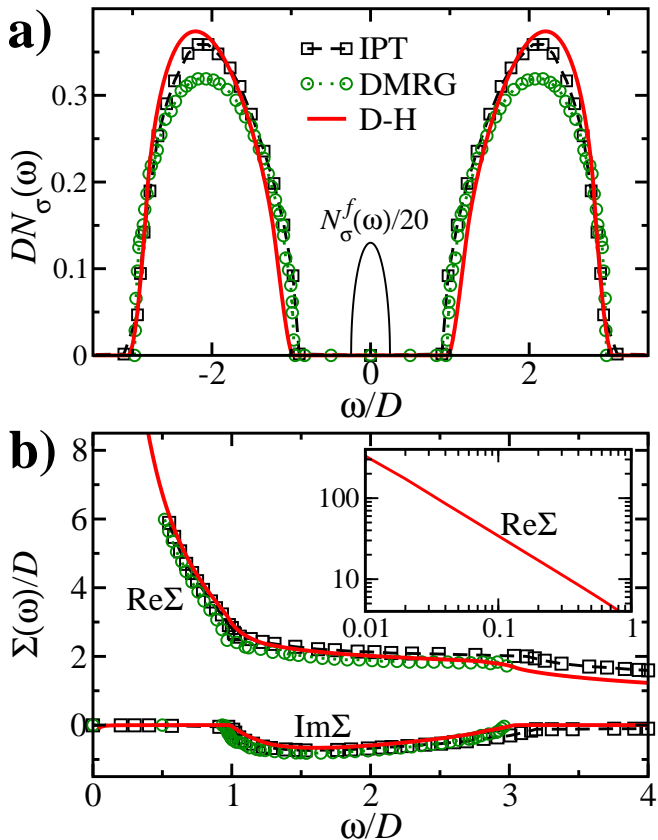

\begin{center}
\fig{3.4in}{fig4.eps}\caption{Spectroscopy of the Mott insulator in the large-$z$ limit of the D-H binding theory (red lines) at $U=4D$.
(a) The spectral density of states. Thin solid line: spinon density of states.
(b) The real and imaginary parts of the electron self-energy. Inset: Real part of self energy on log-log plot, showing the $1/\omega$ dependence.
The DMFT results obtained at $U=4D$  using the zero temperature IPT (open squares) \cite{dmftrmp96} and the dynamical DMRG (open circles) \cite{karski} impurity solvers are also shown for comparison.}
\end{center}
\end{figure}
We note in passing that these properties of the Mott transition and Mott insulator are inaccessible to Gaussian fluctuations around the Kotliar-Ruckenstein saddle point governing the putative BR transition at large $U$ \cite{raimondi,castellani}.

The central quantity in the large-$z$ limit is the electron local self-energy $\Sigma(\omega)$.
It can be extracted by casting the local {\it  electron} Green's function in Eq.~(\ref{convolution}) into the standard form in terms of the self-energy,
\begin{equation}
G_\sigma (\omega)=\int_{-D}^D d\epsilon \rho_0 (\epsilon) {1\over\omega -\epsilon -\Sigma(\omega)}.
\label{gdmft}
\end{equation}
The calculated real part (${\rm Re}\Sigma$) and imaginary part (${\rm Im}\Sigma$) of the {\it  electron} self-energy in the current D-H binding theory are plotted in Fig.~4(b) as a function of $\omega$ at $U=4D$.
For comparison, the DMFT results obtained using the zero temperature IPT and the dynamical DMRG impurity solvers are also shown at the same value of $U=4D$ \cite{dmftrmp96,karski}.
Remarkably close agreement can be seen between both the real and the imaginary part of the self-energies.
Moreover, inside the Mott gap, the real part of the self-energy shows the scaling behavior ${\rm Re}\Sigma(\omega)\propto1/\omega$ as shown on the log-log scale in the inset of Fig.~4(b) in agreement with the DMFT \cite{dmftrmp96}.

\subsection{Gauge field dynamics and deconfinement}
The emergence of the spin-liquid Mott insulator with gapless spinon excitations requires spin-charge separation and is stable only if the gauge field that couples them in Eqs~(\ref{lagrangian}-\ref{hb}) is deconfining.
To derive the gauge field action, we integrate out the matter fields using the hopping expansion \cite{im1995}.
To leading order in $1/z$, the low energy effective gauge field action is given by \cite{longliang},
\begin{eqnarray}\label{action}
S_{\mathrm{eff}}
&=&-\frac{\eta}{z\pi^2}\sum_{\langle i,j\rangle} \int^{\beta}_0 \!\!\mathrm{d} \tau_1 \int^{\beta}_0 \!\!\mathrm{d}\tau_2
\frac{\cos{(a_{ij} -a_{ij}(\tau_2))}} {(\tau_1-\tau_2)^2}\nonumber\\
&&+\frac{1}{zC}\sum_{\langle i,j\rangle}  \int^{\beta}_0 \!\!\mathrm{d} \tau(\partial_{\tau}a_{ij})^2,
\end{eqnarray}
where the second term comes from integrating out the gapped D/H and corresponds to charging with $C$ the ``charging energy'' of a link.
In the large-$U$ limit, $C\propto U^3/t^2$. In the language of the $U(1)$ gauge theory, it describes the electric field action and causes the confinement of the gauge charges.
The first term, which is nonlocal in imaginary time and corresponds to dissipation, comes from the contribution from the gapless fermion spionons.
In our case, $\eta=1$, but we will keep it as a parameter for the dissipation strength in the following discussion.
This term is periodic in the gauge field, reflecting its compact nature.

Thus the gauge field action is dissipative. It has been argued under various settings that a large enough dissipation $\eta$ can drive the compact $U(1)$ gauge field to the deconfinement phase at zero temperature \cite{nagaosa1993, wang2004, ksk2005}.
In the large-$z$ limit, Eq.~(\ref{action}) shows that spatial fluctuations of the link gauge field are suppressed and the dissipative gauge field theory becomes local, i.e. $a_{ij}(\tau) =a(\tau)$.
As a result, the action becomes identical to the dissipative tunneling action derived by Ambegaokar, Eckern, and Sch\"on \cite{aes1982} for a quantum dot coupled to metallic leads, or a shunted Josephson junction with QP tunneling \cite{kampf}.
The $2\pi$-periodicity of the compact gauge field requires $a(\tau)=\tilde a(\tau)+2\pi n \tau/\beta$ where $\tilde a(\tau)$ is single-valued and satisfies $\tilde a(0)=\tilde a(\beta)$, and $n$ is an integer winding number associated with charge quantization, i.e. the instantons in the electric field when charges tunnel in and out of the link.
If the temporal fluctuation of the winding number $n$ is strong, the periodicity of the $a(\tau)$ is important and the gauge field is in the confinement phase.
Otherwise, its compactness is irrelevant and the gauge field is in the deconfinement phase.
For a 2D array of dissipative tunnel junctions, it has been shown that there exists a confinement-deconfinement (C-DC) transition of the winding number at a critical $\eta_c^{2D}\simeq0.45$ \cite{mooji}.
Using the Villain transformation \cite{villain}, one can show that the instanton action is described by a dissipative sine-Gordon model, exhibiting a C-DC transition at a critical dissipation $\eta_c=1/4$.
In our case, $\eta >\eta_c$, and the temporal proliferation of the instantons is suppressed by dissipation \cite{longliang}.
Thus, the gauge electric field is deconfining and the gapless $U(1)$ spin-liquid is indeed the stable Mott insulating state.

\section{Summaries and Discussions}
\begin{figure}
\begin{center}
\fig{3.4in}{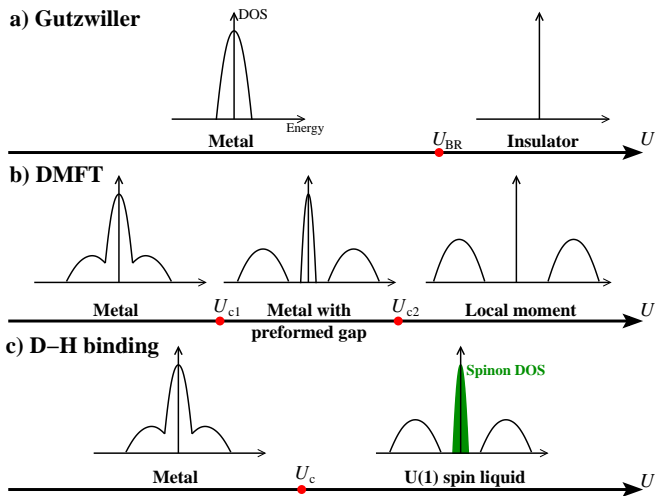}\caption{Schematic diagram of the Mott transition. (a) Gutzwiller with $U_{\rm BR}\simeq3.40D$; (b) DMFT with $U_{c1}\simeq2.38D$ and $U_{c2}\simeq3.04D$ \cite{dmftrmp96,karski,dv2012}; and (c) present D-H binding theory with $U_c\simeq0.8U_{\rm BR}\simeq2.71D$. $D$: half bandwidth.}
\end{center}
\end{figure}

In summary , we have provided an asymptotic solution of the Hubbard model in the large-$z$ limit to capture the most essential Mott physics, i.e., the excitonic binding between oppositely charged doublons and holons \cite{kaplan, yokoyama, capello, leigh, zhouwangwang, mckenzie}.
In the Mott insulator, where the D-H binding theory simplifies considerably as all doublons and holons are bound in real space into excitonic pairs,
the motion of the QP must involve breaking the D/H pairs and thus amounts entirely to incoherent excitations above the charge gap set by the D-H binding energy. We construct a dynamical SBMF theory in the large-$z$ limit and find a continuous Mott transition, where the opening of the Mott gap and the vanishing of the QP coherence coincide at the same $U_c$.
The BR transition is preempted by quantum fluctuations and replaced by the Mott transition.
A key feature of our asymptotic solution is that on the insulating side of the Mott transition, quantum spin fluctuations via the intersite spinon corrlation remain and survive the large-$z$ limit.
The coherent hopping of the spinons gives rise to a gapless QSL by lifting the ground state degeneracy.
The obtained results is in quantitatively agreement with the DMFT with various numerical quantum impurity solvers.
The derived effective action for the compact gauge field in the large-$z$ limit show that the emergent dissipative dynamics drives the gauge field to the deconfinement phase where the spin-charge separated $U(1)$ spin liquid is stable.

To end this paper, we compare the electron spectral function obtained in the present theory to that obtained in other scenarios of the Mott transition.
Focusing exclusively on the coherent QP, Gutzwiller variational wave function approaches \cite{gutzwiller} obtained a strongly correlated Fermi liquid \cite{dv-rmp84} that undergoes a BR transition \cite{br} to a localized state with vanishing QP bandwidth and vanishing doublon D/H density (Fig.~5a).
The single-site DMFT maps the lattice Hubbard model to a quantum impurity embedded in a self-consistent bath \cite{dmftrmp96,dv2012}.
The mapping is exact in the well-defined large-$z$ limit.
The obtained $T=0$ Mott transition shown in Fig.~5b shows that the opening of the Mott gap at $U_{c1}$ and the disappearance of the QP coherence at $U_{c2}$ do not coincide such that the QP states in the metallic state for $U_{c1}<U<U_{c2}$ are separated from the incoherent spectrum by a preformed gap.
This peculiar property \cite{nozieres,kehrein,gebhard} was shown to be correct \cite{kotliar} for the large-$z$ limit taken in the DMFT where the spin-exchange interaction $J\sim t^2/U$ scales with $1/z$ and forces the paramagnetic insulating state to be in a local moment phase with $2^N$-fold degeneracy, i.e., a quantum paramagnet.
In contrast, the current D-H binding theory finds a continuous Mott transition shown in Fig. 5c from a correlated metal to an insulating QSL, where the opening of the Mott gap and the vanishing of the QP coherence coincide at the same $U_c$.

\bigskip

\section{Acknowledgments}

We thank G. Kotliar and Y.P. Wang for useful discussions.
This work is supported by the Key Research Program of Frontier Sicences, CAS, Grant No. QYZDB-SSW-SYS012 (S.Z.), the Strategic Priority Research Program of CAS, Grant No. XDB28000000 (S.Z), the Academy of Finland through its Centres of Excellence Programme (2015-2017) under project No. 284621 (L.L.), and the U.S. Department of Energy, Basic Energy Sciences Grant No. DE-FG02-99ER45747 (Z.W.). Z.W. thanks the hospitality of Aspen Center for Physics (ACP) where this work was conceived, and the support of ACP NSF grant PHY-1066293. Numerical calculations were performed on the HPC Cluster of ITP-CAS.

\end{document}